%% file: higgs3.tex
\documentclass[11pt]{article}
\usepackage{epsfig,sint,macros}
\begin{document}
\input title.tex
\input sect1.tex
\input sect2.tex
\input sect3.tex
\input sect4.tex
\input concl.tex
%
   \bibliography{lattice}        
   \bibliographystyle{h-elsevier}   
\end{document}

%% file: title.tex
\begin{titlepage}
\begin{flushright}
  COLO-HEP-444 \\
  DESY 00-078
\end{flushright}

\vskip 1 cm
\begin{center}
  {\Large\bf String Breaking as a Mixing Phenomenon in the SU(2) Higgs Model }
\end{center}
\vskip 1 cm
\vbox{
\centerline{
\epsfxsize=2.5 true cm
\epsfbox{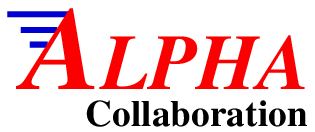}}
}
\vskip 1.0cm
\begin{center}
{\large Francesco Knechtli $^{a~\dagger}$ and Rainer Sommer $^{b~\ddagger}$}
\vskip 0.8cm
$^{a}$ Physics Department, 
                     University of Colorado, Boulder, CO 80309 USA
\vskip 0.4cm
$^{b}$ DESY, Platanenallee 6, D-15738 Zeuthen, Germany
\end{center}
\vskip 2.5ex
{\bf Abstract}
\vskip 0.7ex
We study the potential of a static quark anti-quark pair in the
confinement ``phase'' of the SU(2) Higgs model. Around separation $r_b$,
the confining string of the gauge field breaks by formation of a
dynamical pair of light quarks. 
The ground state and first
excited state static potentials are determined by a variational
technique from a
matrix correlation in which suitably smeared gauge and Higgs fields enter. 
Our results at $\beta=2.4$ clearly show string breaking ($r_b\approx1.9\rnod$).
The investigation of properly
defined overlaps confirms the interpretation of string breaking as a 
level crossing
phenomenon between string-type and meson-type states.
We study the scaling properties of the static potentials along
a line of constant physics, varying the lattice spacing by a 
factor of 2. Our results show compatibility with scaling within tiny errors.
  \vfill

\begin{flushleft}
  COLO-HEP-444 \\
  DESY 00-078 \\
  May 2000
\end{flushleft} 
$^{\dagger}$ {\small e-mail: knechtli@pizero.colorado.edu} \\
$^{\ddagger}$ {\small e-mail: sommer@ifh.de}
\eject

\vfill

\eject

\end{titlepage}


%% file: sect1.tex
\section{Introduction \label{s_intro}}

The potential of a static quark\footnote{
A static quark (anti-quark) is an external source in the (complex
conjugate of the) fundamental representation of the gauge group.}
anti-quark pair in a Yang-Mills gauge theory has been computed by
Monte Carlo simulations on the lattice. The observable from which the
static potential is extracted is the Wilson loop.
The results for gauge group SU(2)
\cite{Booth:1993bk,Sommer:1993ce} and SU(3)
\cite{Bali:1992ab,Booth:1992bm}, close to the continuum limit, show a
linearly rising confinement potential at large separations of the
static charges.
When the Yang-Mills gauge theories are coupled to matter fields in the
fundamental representation of the gauge group,
the static potential is expected to
flatten at large distances: the ground state of the system is better
interpreted in terms of two weakly interacting static-light mesons which are
bound states of a static and a dynamical quark.
The dynamical quarks are pair-created in the strong
gauge field binding the static quarks. 
This phenomenon is
called {\em string breaking} or {\em screening} of the static
charges. The name ``string'' refers to the gauge 
field configuration which confines the static quarks and leads to the 
linear confinement in pure gauge theories. 

In recent attempts in QCD with two flavors of dynamical quarks 
\cite{Glassner:1996xi,Gusken:1997sa,Allton:1998gi,Burkhalter:1998wu,Aoki:1999ff,Schilling:Pisa99,Bolder:2000un}, 
the flattening in the static potential determined
from the Wilson loops was not visible.
The string breaking distance $\rb$ around which the static potential
should start flattening off, could nevertheless be estimated in the
quenched approximation of QCD to be \cite{Alexandrou:1994ti,Sommer:1996fr} 
\bes\label{rbquenched}
 \rb & \approx & 2.7\rnod \,,
\ees
where $\rnod\approx0.5\,\fm$ is a scale conveniently defined from the
force, $F(r)$, between static quarks\cite{Sommer:1993ce}:
\bes
 F(\rnod)\rnod^2 & = & 1.65\rnod \,,
\ees
The QCD results so far show a linear rise of the static potential for distances
beyond $\rb$. It has been commonly argued that the problem is the 
poor overlap of the Wilson loops with the ground state of the system.
The investigation of the static potential in models other than QCD is
therefore relevant in order to understand its origin
and identify the reason for the failure of the method
used to extract it in full QCD.

First studies of string breaking were performed with a
hopping-parameter expansion in SU(2) gauge theory with Wilson fermions
\cite{Joos:1983qb}. In the Schwinger model ($\rm{QED}_2$),
the exact solution for the static potential
can be given in the limit of zero fermion mass \cite{Becher:1983ft}: 
$V(r)=(e\sqrt{\pi}/2)\{1-\exp(-er/\sqrt{\pi})\}$, where
$e$ is the charge of the static sources. String breaking was
established by numerical simulation in the Schwinger model
\cite{Potvin:1985gw,Dilger:1992yn}.
Numerical evidence of the screening of the static potential was also
found in the U(1) Higgs model (scalar QED) in two dimensions
\cite{Jochen:PhD}.
The flattening of the static potential
at large distances is also expected in the confinement ``phase'' of the
SU(2) Higgs model.
Indeed, early simulations yielded some
qualitative evidence for string breaking
\cite{Evertz:1986vp,Bock:1990kq}.

String breaking can also be studied in Yang-Mills theories using
static sources in the adjoint representation of the gauge group. The
gauge field itself is responsible for the screening of the
sources and the formation of hadrons called ``gluelumps''. 
An important numerical investigation concerning this screening has
been carried out by
C. Michael in \cite{Michael:1992nc}, where it has been
noted that string breaking can be a {\em mixing}
phenomenon. The static potential is extracted from a matrix
correlation in which two types of states enter,
the adjoint string and the ``two-gluelump''. 
However, due to large errors, no clear evidence for
string breaking could be given.
The first numerical evidence for string
breaking in non-Abelian gauge theories with dynamical matter fields
was given using the mixing method in the four-dimensional \cite{Knechtli:1998gf} and
three-dimensional \cite{Philipsen:1998de} SU(2) Higgs model by the
computation of the potential between static quarks.
Most recently, the
extraction of the static adjoint potential in the three-dimensional
\cite{Stephenson:1999kh,Philipsen:1999wf} and four-dimensional
\cite{deForcrand:1999kr} SU(2) Yang-Mills theory
shows also evidence for string breaking.

In full QCD, string breaking has been seen at finite
temperature \cite{DeTar:1998qa}, where the static potential can be
extracted from Polyakov loop correlators. A recent investigation
in zero-temperature QCD with two flavors of dynamical quarks
\cite{Pennanen:2000yk} gives some indication that string 
breaking can be observed in
the potential determined from a matrix correlation containing string-type
and meson-type states. The main problems are the
computational costs of the light quark propagators entering the matrix
correlation: the maximal variance reduction method of
\cite{Michael:1998sg} has been applied in \cite{Pennanen:2000yk}.

In this article, we present results for the spectrum of static-light mesons
and for the static potential in the confinement
``phase'' of the SU(2) Higgs model. The method of computation of the
static potential is the same as in our first work
\cite{Knechtli:1998gf} but it is explained here in much more detail. 
In \sect{s_sl_mesons},
we present a detailed investigation of the
spectrum of the static-light mesons. This is relevant not only for
determining the asymptotic value of the static potential, but also for
finding a suitable smearing procedure for the Higgs field to  be
used in the computation of the static potential.
In \sect{s_sb}, we show
our results for the ground state and the first excited static
potential obtained at $\beta=2.4$, with a spatial resolution two times
better than at $\beta=2.2$ used in \cite{Knechtli:1998gf}.

In order to study the interpretation of string breaking as a mixing 
phenomenon
\cite{Drummond:1998ar} we first properly define overlaps of
the string and two-meson states with the ground and first
excited energy eigenstates. Their dependence on the distance $r$ then 
establishes string breaking as a level crossing phenomenon. 
In \sect{s_scaling}, we study the scaling of the static
potentials by comparing the results at $\beta=2.4$ with $\beta=2.2$
on a line of constant physics \cite{Knechtli:1999qe}.
The dependence of the static potentials on the value of
the Higgs quartic self-coupling is also investigated.


%% file: sect2.tex
\section{The spectrum of the static-light mesons \label{s_sl_mesons}}

The investigation of the spectrum of static-light mesons is an important
study to be done before the extraction of the static potential. The
system composed by a static quark anti-quark
pair is expected to be described at large separation of the static sources
by two weakly interacting static-light
mesons, which are bound states of a static quark and the dynamical
matter field. Denoting by $\mu$ the mass of the lowest meson state, the
static potential $V_0(r)$ is expected to approach the value
\bes\label{potasympt}
 \lim_{r\to \infty}V_0(r) = 2 \mu \,.
\ees
Therefore, the mass $\mu$ basically determines the string
breaking distance $\rb$ around which the potential starts flattening out. 

The extraction of the static-light meson spectrum is representative for
the variational method that we employ also for the extraction of the
static potential.
We constructed a large basis of operators that create one-meson
``states'' when applied to the vacuum. The variational approach chooses the
best linear combinations of these ``states'' which approximate the energy
eigenstates. Through the determination of the meson spectrum we can
therefore gain information about a suitable way of constructing
the meson-type states. Then, we use this information in \sect{s_sb} to
construct a basis of states for the determination of the static potentials.

\subsection{The matrix correlation}

In \cite{Francesco:PhD}, an Hamiltonian formalism for the SU(2) Higgs
model is constructed. Along the lines of \cite{Luscher:TM}, a {\em transfer
matrix operator} is defined and its strict positivity for $\kappa>0$ and
$\lambda>0$ is proved\footnote{
One can show that
the partition function of the SU(2) Higgs model satisfies the property
$Z(\beta,\kappa,\lambda)=Z(\beta,-\kappa,\lambda)$.
There is a mapping of the observables such that the expectation values at
positive $\kappa$ are reproduced by expectation values at negative
$\kappa$. This motivates the restriction of the parameter region to the values
$\kappa>0$. For $\lambda=0$, strict positivity of the transfer matrix
holds for $0<\kappa<1/6$.
}.
This property is equivalent to the reality of the energy spectrum in
any sector of the Hilbert space. Different {\em charge sectors} of the Hilbert 
space are
defined by the transformation property of the states under gauge
transformation. Through Gauss' law, this gauge transformation property is related
to the presence of static charges in some irreducible representation
of the gauge group. 

The static-light meson states belong to the charge sector with one static charge
in the fundamental representation of the gauge group localised at a certain
space position $\vec{x}$. The meson states $|i\rangle,\,(i=1,2,3,...),$
are described by
(composite) fields $O^{\rmM}_i(x)$, which are constructed with field
variables taken at equal time $x_0$ and transform under a gauge
transformation $\{\Lambda(x)\in\SUtwo\}$ according to
\bes\label{gtrsfmesonfield}
 [O_i^{\rmM,\Lambda}(x)]_a & = &
 \Lambda_{aa^{\prime}}^{\dagger}(x)[O_i^{\rmM}(x)]_{a^{\prime}} \,,
\ees
where $a,\,a^{\prime}=1,2$ are color indices. The obvious choice is to
take $O^{\rmM}(x)$ to be the Higgs field $\Phi(x)$. We are also going to
consider non-local linear combinations which take into account contributions from
Higgs fields at neighboring sites (smeared fields) and more
general composite fields, with the intent to model the true  wave function
of the meson. From the basis of meson-type fields
$O^{\rmM}_i(x)$ a matrix correlation
\bes\label{mucorr}
  C^{\rm M}_{ij}(t) & = & \langle [O^{\rmM}_j(x+t\hat 0)^*]_a\,
  U(x,x+t\hat 0)^{\dagger}_{ab}\,[O^{\rmM}_i(x)]_b \rangle \,.
\ees
is constructed representing the transition amplitude over a time
interval $t$ from the meson state $i$ to the meson state $j$.
The static charge is represented by a straight time-like Wilson line
$U(x,x+t\hat 0)^{\dagger}$ connecting $x$ with $x+t\hat{0}$. With the
help of the reconstruction theorem proved in \cite{Francesco:PhD}, it is
possible to show that in the limit of an infinite physical time
extension $T$ of the lattice\footnote{
In practice, the limit $T\to\infty$ is reached when
$Tm_{\rmH}\gg1$, where $m_{\rmH}$ is the Higgs mass defined as
the mass gap in the zero charge (gauge invariant) sector of the Hilbert
space.}
the correlation matrix \eq{mucorr} can be written like
\bes\label{mucorrtd}
 C^{\rm M}_{ij}(t) & = & \sum_{\alpha} 
 \langle j|\alpha\rangle 
 \langle \alpha|i \rangle\,\rme^{-tW_{\alpha}} \,,
\ees
where $|\alpha\rangle,\,(\alpha=0,1,2,...)$ are the orthonormal meson
energy-eigenstates with energies\footnote{
We normalise the vacuum energy to be 0.}
$W_{\alpha}$, $W_{\alpha}<W_{\alpha+1}$.
The matrix correlation
\eq{mucorr} can be measured in a Monte Carlo simulation: now, we
describe the variational method for extracting the meson energy
spectrum from it.

\subsection{Variational method \label{variation}}

For matrices of the type in \eq{mucorrtd} a general lemma for the
extraction of the energies $W_{\alpha}$ has been proved in
\cite{phaseshifts:LW}. In this reference, a variational
method is proposed, which is superior to a straightforward application
of the lemma. It consists in solving the generalised
eigenvalue problem:
\bes\label{genev}
  \sum_j C_{ij}(t)v_{\alpha,j}(t,t_0) & = & 
  \lambda_{\alpha}(t,t_0)\sum_j C_{ij}(t_0)v_{\alpha,j}(t,t_0) \, , \quad 
  \lambda_{\alpha} > \lambda_{\alpha+1} \, ,
\ees
where $t_0$ is fixed and small (in practice we use $t_0=0$).
The generalised eigenvalues
$\lambda_{\alpha}(t,t_0)$ are computed as the
eigenvalues of $\bar{C}=C(t_0)^{-1/2}C(t)C(t_0)^{-1/2}$
and the vectors
\bes\label{genevec}
 \bar{v}_{\alpha,i}\;=\;\sum_j [C(t_0)^{1/2}]_{ij}v_{\alpha,j}(t,t_0) 
 & \mbox{with} &
 \sum_i \bar{v}_{\alpha,i}\bar{v}_{\alpha^{\prime},i}\;=\;
 \delta_{\alpha\alpha^{\prime}}
\ees
are the orthonormal eigenvectors of $\bar{C}$.
The positivity of the transfer matrix
ensures that $C(t)$ is positive definite for all $t$.
In \cite{phaseshifts:LW} it is proven that the energies
$W_{\alpha}$ are given by the expressions
\bes\label{spectrum}
  a W_{\alpha} & = & \ln(\lambda_{\alpha}(t-a,t_0) 
  /\lambda_{\alpha}(t,t_0)) +
  \rmO\left(\rme^{-t\Delta W_{\alpha}}\right) \, ,
\ees
where $\Delta W_{\alpha}=
\min\limits_{\beta\neq\alpha}|W_{\alpha}-W_{\beta}|$. It is
expected that, for a good basis of states,
the coefficients of the higher exponential corrections
in \eq{spectrum} are suppressed so that the
energies can be read off at moderately large values of $t$ from the
right-hand side of \eq{spectrum}.

The variational method \eq{genev} and \eq{spectrum} is our standard
method for extracting the energy spectrum. What we have stated here
about this method is valid for
{\em any} charge sector of the Hilbert space. One has to start from a
basis $|i\rangle$ of states belonging to that charge sector. The matrix
correlation $C_{ij}(t)$ corresponds to matrix elements
$\langle j|\trans^n|i\rangle,\;n\equiv t/a,$
of powers of the transfer matrix operator $\trans$
appropriate for the charge sector. In \cite{Francesco:PhD}, this
correspondence is derived in detail for the sector with a static quark
anti-quark pair: the energy spectrum are the static potential and its
excitations.

\subsection{Smeared Higgs fields}

We studied different bases of meson-type fields $O_i^{\rmM}(x)$
by measuring in Monte Carlo simulations the matrix correlation function
$C_{ij}(t)$ defined in \eq{mucorr} and computing from it the energy
spectrum of the static-light mesons using the variational method described
in \sect{variation}.
Our aim was to find the best field basis for describing
the ground state of the static-light mesons.
For these studies we simulated the SU(2) Higgs model on a $20^4$ lattice
with parameters $\beta=2.2$, $\kappa=0.274$ and $\lambda=0.5$. This
parameter point is in the confinement ``phase'' of the model and is the
point that we used in our first work \cite{Knechtli:1998gf}.
The measurement of the matrix correlation is improved by the use of the
one-link integral method \cite{Parisi:1983hm}.
\begin{figure}[tb]
\hspace{0cm}
\vspace{-1.0cm}
\centerline{\psfig{file=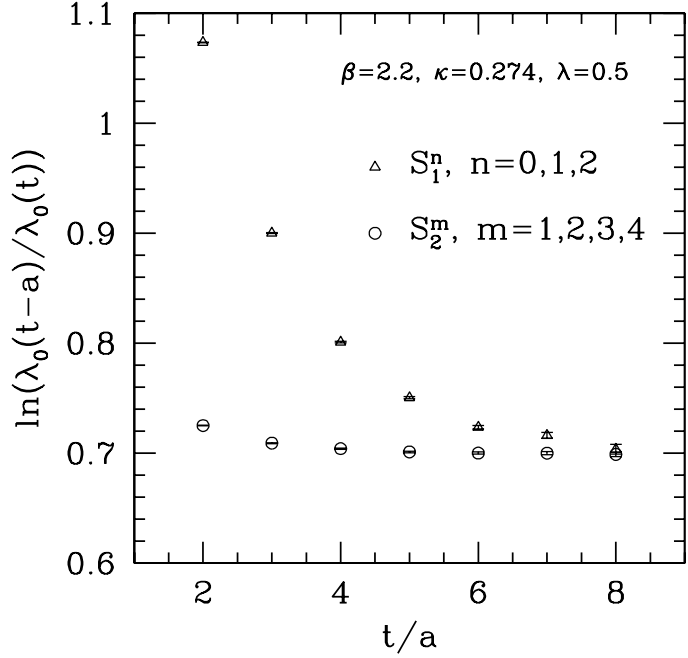,width=10cm}}
\vspace{-0.0cm}
\caption{Here, we compare the extraction of the mass $\mu$ of a
  static-light meson using different smearing operators defined in
  \eq{smearophiggs1} and \eq{smearophiggs2}. The smearing levels used are
  indicated by $n$ and $m$.
\label{meson_comp}}
\end{figure}

We first studied a basis containing the fundamental Higgs field
$\Phi(x)$ and smeared Higgs fields obtained by iterating the
application of a smearing operator $S_1$ to the Higgs field.
The smearing operator $S_1$ is defined as
\bes\label{smearophiggs1}
  S_1\,\Phi(x) & = &  
  \Phi(x)+\sum_{|x-y|=a \atop x_0=y_0}U(x,y)\Phi(y) \, , 
\ees
where $U(x,y)$ is the link connecting $y$ with $x$. Iterating the
smearing operator $S_1$ we obtain smeared Higgs fields
$\Phi^{(m)}_1(x) = S_1^m\,\Phi(x),\,(m=0,1,2,...),$ 
with different smearing levels $m$
($m=0$ corresponds to the fundamental Higgs field in the Lagrangian).
We measured a matrix correlation function with a
basis of smeared Higgs fields corresponding to smearing
levels 0,1 and 2 of $S_1$. The result for the ground state extracted
according to \eq{spectrum} is shown in \fig{meson_comp} (triangles). 
We were not able to reach a plateau for the ratio 
$\ln(\lambda_{\alpha}(t-a)/\lambda_{\alpha}(t))$ within the
range of $t$ considered (up to 8 in lattice unit).
\begin{figure}[tb]
\hspace{0cm}
\vspace{-1.0cm}
\centerline{\psfig{file=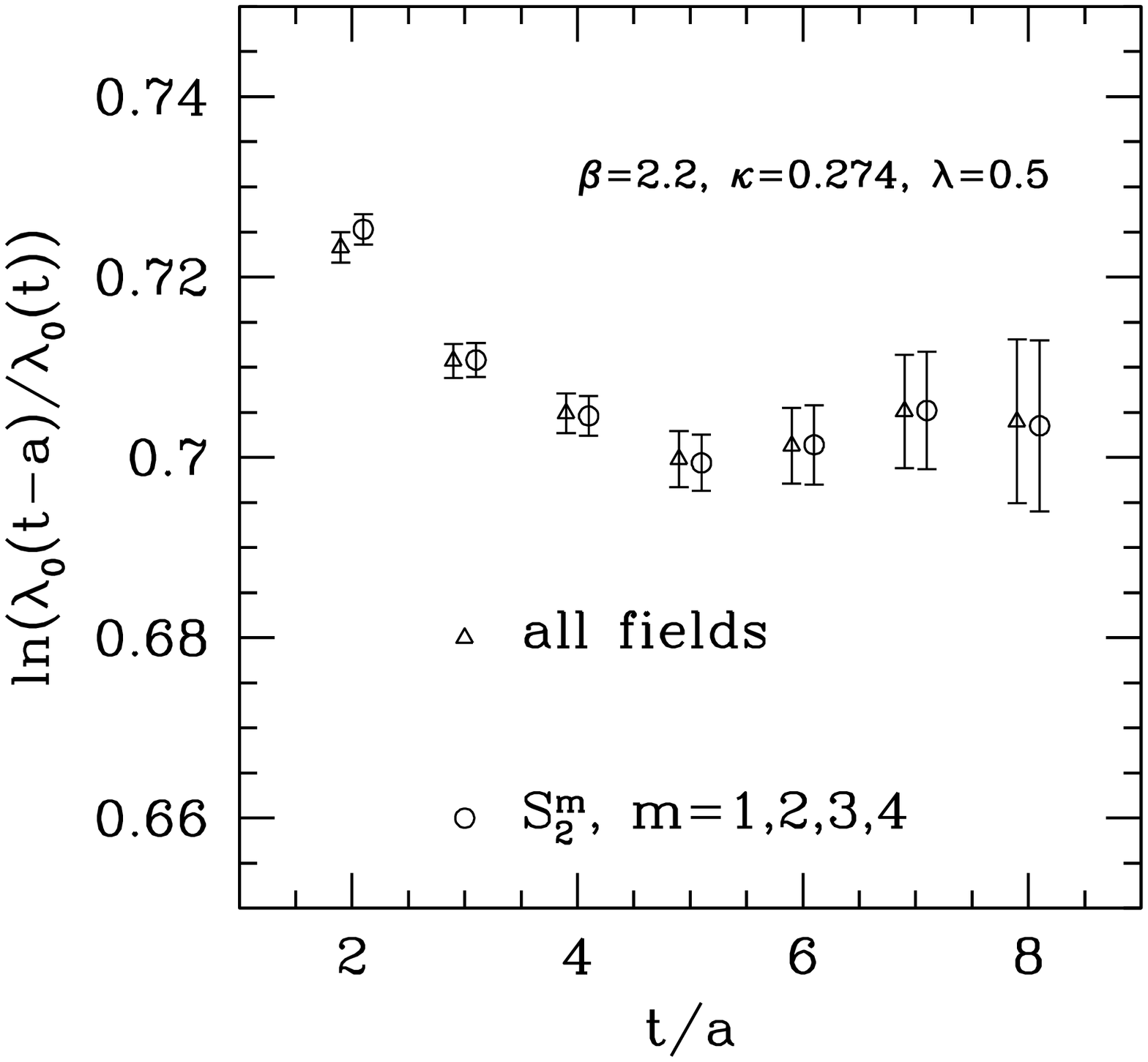,width=10cm}}
\vspace{-0.0cm}
\caption{Here, we compare the extraction of the mass $\mu$ of a
  static-light meson using all the fields \eq{meop1}--\eq{meop2}
  and only the smeared Higgs fields corresponding to smearing levels
  1,2,3,4 of the smearing operator $S_2$ defined in \eq{smearophiggs2}.
\label{meson_all}}
\end{figure}

Then, we investigated a larger basis of meson-type fields, defining in
particular a smearing operator $S_2$ as
\bes\label{smearophiggs2}
  S_2\,\Phi(x) & = &
  \mathcal{P}\{\mathcal{P}\Phi(x) + 
  \mathcal{P}\sum_{|x-y|=\sqrt{2}a \atop x_0=y_0}
  \overline{U}(x,y)\Phi(y) + \nonumber \\ 
  & & \quad\, \mathcal{P}\sum_{|x-y|=\sqrt{3}a \atop x_0=y_0}
  \overline{U}(x,y)\Phi(y)\} \, , 
\ees
where $\mathcal{P}\Phi = \Phi/\sqrt{\Phi^{\dagger}\Phi}$
and  $\overline{U}(x,y)$ represents
the average over the shortest link connections between $y$ and $x$.
Through iteration of $S_2$ we obtain the smeared Higgs fields
$\Phi^{(m)}_2(x) = S_2^m\,\Phi(x),\,(m=0,1,2,...)$.
We considered the following
basis of meson-type fields $O^{\rmM}_i(x),\; i=1,2,...,11$:
\bes
 O^{\rmM}_1(x) & = & \mathcal{P}\Phi(x) \, , \label{meop1} \\
 O^{\rmM}_2(x) & = & \mathcal{P}\sum_{|x-y|=a \atop x_0=y_0} 
  U(x,y)\Phi(y) \, , \\
 O^{\rmM}_3(x) & = & \mathcal{P}\sum_{|x-y|=\sqrt{2}a \atop x_0=y_0}
  \overline{U}(x,y)\Phi(y) \, , \\
 O^{\rmM}_4(x) & = & \mathcal{P}\sum_{|x-y|=\sqrt{3}a \atop x_0=y_0}
  \overline{U}(x,y)\Phi(y) \, , \\
 O^{\rmM}_i(x) & = & \Phi^{(i-4)}_2(x)\,, \quad i=5,6,7,8 \, , \\
 O^{\rmM}_9(x) & = & \Phi(x)\times\frac{1}{6}\sum_{k=1}^3\{
  \Phi^{\dagger}(x-a\kh)U(x-a\kh,k)\Phi(x)+ \nonumber \\
 & & \qquad\qquad\qquad \Phi^{\dagger}(x)U(x,k)\Phi(x+a\kh) \}
  \, , \\
 O^{\rmM}_{10}(x) & = & \Phi(x)\times\frac{1}{12}\sum_{1\leq k<l\leq3}
  \{P_{kl}(x) + P_{kl}(x-a\kh) + \nonumber \\
 & & \qquad\qquad\qquad  P_{kl}(x-a\kh-a\hat{l}) + 
  P_{kl}(x-a\hat{l})\} \, , \label{meplaq} \\
 O^{\rmM}_{11}(x) & = & \Phi(x)\times(\Phi^{\dagger}(x)\Phi(x)) \, ,
  \label{meop2} 
\ees
where in \eq{meplaq}
$P_{kl}(x)=U(x,k)U(x+a\kh,l)U^{\dagger}(x+a\hat{l},k)U^{\dagger}(x,l)$ and we use the
same notation conventions as in \cite{Knechtli:1998gf}.
In \fig{meson_all}, the result for the extraction of the mass of a
static-light meson using the fields $O^{\rmM}_i(x),\;i=1,...,11$ is shown
(triangles). Note the enlarged scale on the y-axis as compared to
\fig{meson_comp}.
We obtain a nice plateau already at moderately large values of
$t$. The situation remains practically unchanged (also the statistical
errors) if we remove from the basis all fields except the smeared fields
obtained by iterations of the smearing operator $S_2$. This means that
this smearing procedure contains all relevant features for describing
the ground state which could be obtained by using the larger
basis.

When the generalised eigenvalue problem \eq{genev} is solved, the
optimal linear combination of the basis fields $O_i^{\rmM}(x)$
describing the ground state can be expressed in terms of the
components of the vector $v_0$ as $\sum_iv_{0,i}O_i^{\rmM}(x)$.
Therefore, we call $v_0$ the ground state wave function.
An interesting fact we can learn from
$v_0$, is that the field $O^{\rmM}_2$, with nearest neighbor
contributions, has a very small coefficient $v_{0,2}$. This explains our
original difficulties in extracting the meson ground state. In
\fig{meson_comp}, a direct comparison of the smearing operators $S_1$
and $S_2$ shows clearly that
the contributions from the excited states are much more suppressed when
we use $S_2$ (circles).
\begin{figure}[tb]
\hspace{0cm}
\vspace{-1.0cm}
\centerline{\psfig{file=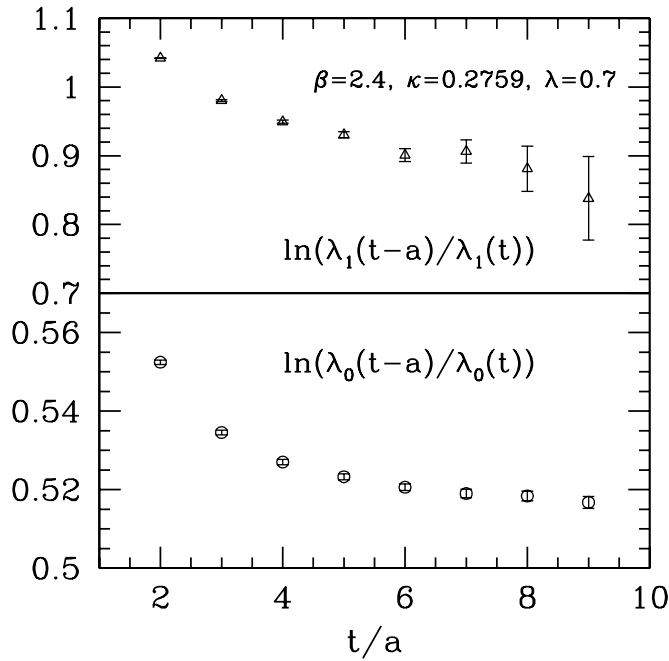,width=10cm}}
\vspace{-0.0cm}
\caption{Here, we show the extraction of the mass of the ground 
  and first excited meson state at $\beta=2.4$. 
  The basis of meson-type fields was obtained using the smearing
  procedure $S_2$ \eq{smearophiggs2} with smearing levels $m=1,3,5,7,10,15$.
  The simulation was performed on a $32^4$ lattice and the statistics
  is of 800 measurements.
\label{meson}}
\end{figure}

\subsection{The meson spectrum at ${\bf \beta=2.4}$}
\begin{figure}[tb]
\hspace{0cm}
\vspace{-1.0cm}
\centerline{\psfig{file=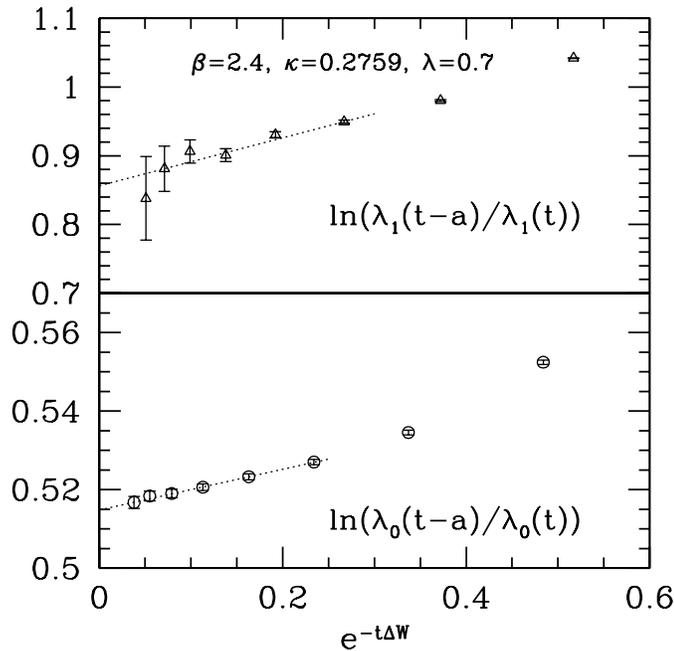,width=10cm}}
\vspace{-0.0cm}
\caption{Here, we show the same data as in \fig{meson} but
  plotted against the correction term $\exp(-t\Delta W)$ in
  \eq{spectrum}. We use $\Delta W=\mu^*-\mu$ for the ground
  state and $\Delta W=\mu^{**}-\mu^*$ for the first excited state, see
  (\ref{muspectrumb24}). \label{f_meson_exp}}
\end{figure}

In \fig{meson}, we show the results for the static-light meson spectrum
that we obtained for the parameters
$\beta=2.4,\;\kappa=0.2759,\;\lambda=0.7$ (in the confinement
``phase'') on a $32^4$ lattice. More details
about this simulation will be given in \sect{s_sb}. For the
measurement of the matrix correlation function we used a basis with
the six fields
\bes\label{basisme24}
 \Phi_2^{(m)}(x)\,, \quad m=1,3,5,7,10,15 \,,
\ees
obtained by iterating the smearing procedure $S_2$ in
\eq{smearophiggs2}. As we will see in
\sect{s_sb}, the lattice spacing at $\beta=2.4$ is reduced by
almost a factor two with respect to the lattice spacing at
$\beta=2.2$. Therefore, at $\beta=2.4$ smeared fields with
high smearing levels $m$
are expected to play a more important role than at $\beta=2.2$. This
expectation is confirmed by the simulation.
In order to determine with confidence the static-light meson masses,
we plot in \fig{f_meson_exp} the logarithmic ratios on the right-hand
side of \eq{spectrum} as
functions of the correction terms $\exp(-t\Delta W)$. This enables us to
choose the best time $t$ for reading off the masses from the logarithmic
ratios and to estimate the systematic errors associated with this
choice. For the mass of the ground state,
we must take the largest value $t/a=9$.
For the mass of the first excited state, we
can take $t/a=8$. In both cases, the systematic errors\footnote{
The systematic errors for the masses are estimated from the
difference between the mass read off at the chosen value of $t$ and
the crossing point of the dotted lines in \fig{f_meson_exp}
with the y-axis ($t=\infty$).}
are of the same magnitude as the statistical errors. However, these
errors are small.  The results for the meson spectrum are\footnote{
We use the notation $\mu$, $\mu^*$ and $\mu^{**}$ for $W_0$, $W_1$ and
$W_2$ respectively.}
\bes\label{muspectrumb24}
 a\mu\;=\;0.517(2)\,, & a\mu^*\;=\;0.88(3)\,, & a\mu^{**}\;=\;1.21(9) \,.
\ees
We note that the convergence of the right-hand
side of \eq{spectrum} is not so ``critical'' in the case of the static
potentials considered in \sect{s_sb}.


%% file: sect3.tex
\section{String breaking and mixing\label{s_sb}}

As mentioned in the introduction, the basic point concerning the determination
of the static potential has been first noted by C. Michael
\cite{Michael:1992nc} in a study of the SU(2) static adjoint potential.
The energy-eigenstates of the system composed of a
static quark anti-quark pair and of light dynamical matter fields
are well described by a superposition of string-type and meson-type states.
Using a suitable basis of such states a matrix correlation can be constructed
from which the static potential and its excitations
are extracted for arbitrary separations of the static quarks.

\subsection{The matrix correlation}

The static potentials are defined as the energy levels in the
charge sector of the Hilbert space with a static quark
at space position $\vec{x}$ and a static anti-quark at space position
$\vec{x}_r=\vec{x}+r\hat{k}$. States living in this charge sector
are described by fields $O_{ab}(x,x_r)$ with color indices $a,b$
and equal times $x_0=y_0$, which transform under
gauge transformations like \cite{Francesco:PhD,Rainer:PhD}
\bes
 O^{\Lambda}_{ab}(x,x_r) & = &
 \Lambda^{\dagger}_{aa^{\prime}}(x)\,O_{a^{\prime}b^{\prime}}(x,x_r)\,\Lambda_{b^{\prime}b}(x_r) \,.
\ees
The simplest choice of such fields describing string-type states
is $U_{ab}(x,x_r)$ and for the meson-type states $\Phi_a(x)\Phi^{\dagger}_b(x_r)$. 
By $U(x,y)$ we denote the product of gauge links along the straight line
connecting $y$ with $x$.
The basic matrix correlation for the extraction of the
static potentials can be expressed in terms of the following
transition matrix elements \cite{Knechtli:1998gf}
\bes
 C_{\rm WW}(r,t) & = & 
  \langle \tr[U(x,x_r)\, 
  U(x_r,x_r+t\hat{0})\,U^{\dagger}(x+t\hat{0},x_r+t\hat{0})\,
  U^{\dagger}(x,x+t\hat{0})] \rangle \,, \label{cww} \\
 C_{\rm WM}(r,t) & = &
  \langle \Phi^{\dagger}(x+t\hat 0)\,
  U^{\dagger}(x,x+t\hat 0)\,U(x,x_r) 
  \, U(x_r,x_r+t\hat 0)\,\Phi(x_r+t\hat 0) \rangle \,, \label{cwm} \\
 C_{\rm MM}(r,t) & = &
  \langle \Phi^{\dagger}(x+t\hat 0)
  U^{\dagger}(x,x+t\hat 0)\Phi(x) \,\,\, 
  \Phi^{\dagger}(x_r)U(x_r,x_r+t\hat 0)\Phi(x_r+t\hat 0)
  \rangle \,. \label{cmm}
\ees
The static quark (anti-quark) is represented by a straight time-like Wilson line
$U^{\dagger}(x,x+t\hat{0})$ ($U(x_r,x_r+t\hat{0})$).
The matrix $C$ is real, symmetric and positive.
This simplest choice of the states does not however correspond to the physical picture
that we have of the system.

The string-type states should reproduce the flux tube
\cite{Bali:1995de,Luscher:1981iy,Sommer:1987uz,Wosiek:1987kx,Caselle:1996fh,Pennanen:1997qm}
of the gauge field binding the static quarks.
Therefore, we smear the space-like links describing the string-type
states using the APE smearing procedure of \cite{smear:ape}
with smearing strength set to the numerical value $\epsilon=1/4$.
In order to construct ``physical'' two-meson states,
we determine the spectrum of the static-light mesons
as described in \sect{s_sl_mesons}.
In the matrix correlation \eq{mucorr}
we use smeared Higgs fields $O^{\rmM}_i(x)=\Phi_2^{(n_i)}(x)$ defined with the 
smearing operator of \eq{smearophiggs2}.
The numbers $n_i\;(i=1,2,...,N)$ denote the smearing levels.
The eigenvectors $v_{\alpha}\in\blackboardrrm^N\;(\alpha=0,1,2,...)$,
obtained by solving
the generalised eigenvalue problem \eq{genev} for large $t$,
are the wave functions
describing approximately (because of the finite basis of fields and 
the finite time $t$) the true eigenstates of the Hamiltonian.
We define the fields
\bes\label{mesoneig}
 \Psi_{\alpha}(x) & = & 
 \sum_{i=1}^N v_{\alpha,i}\Phi_2^{(n_i)}(x)  \quad
 (\alpha=0,1,2,...) \,,
\ees
corresponding to the approximate one-meson eigenstates. The fields we
choose to describe two-meson states are defined as
\bes\label{twomeson}
 \left[\Psi_{\alpha}(x)\right]_a\cdot
 \left[\Psi_{\beta}^*(x_r)\right]_b \, , \quad
 \alpha,\beta\,=\,0,1,2 \,.
\ees
The values $\alpha=0,1,2$ refer to the ground,
first and second excited one-meson state. The field basis in \eq{twomeson}
contains combinations with $\alpha\neq\beta$ 
which are not symmetric under interchange of the positions $x$ and
$x_r$ of the static charges.
Because we expect the ground two-meson state to be symmetric,
we project into the symmetric 
linear combinations of the fields in
\eq{twomeson} when we analyse the data of the simulations. The ``mixed''
states (for example of one meson in the ground state and one meson in
the first excited state) can be important when looking at the 
asymptotic behavior (in $r$) of excited static potentials
\cite{prniedermayer}.
The one-meson states have
a space extension due to the smearing of the Higgs
field. For a high number of smearing iterations, there is
effectively an ``interaction'' between the mesons in the two-meson
states \eq{twomeson} due to the overlap of the smeared Higgs fields

Summarising, we use the basis of states $|i\rangle$ 
described by the fields
\bes\label{potbasis}
 [O_i(x,x_r)]_{ab} =  
 \left\{\bea{cl} U^{(m_i)}_{ab}(x,x_r) &
 i=1,2,...,N_{\rmU} \\
 \left[\Psi_{\alpha_i}(x)\right]_a\,\left[\Psi_{\beta_i}^*(x_r)\right]_b &
 i=N_{\rmU}+1,...,N_{\rmU}+9 \ea \right.
\ees
where $U^{(m_i)}(x,x_r)$ is the product of smeared gauge links
(with smearing level $m_i$) along the straight line connecting $x_r$
with $x$ and the pairs of indices
$(\alpha_i=0,1,2;\beta_i=0,1,2)$ label the 9 combinations of
two-meson states. We construct the following matrix correlation
\bes\label{potcorrev}
 C_{ij}(t,r)\,=\,\langle [O_i(x,x_r)]_{ab} \, U_{bc}(x_r,x_r+t\hat{0}) \,
 [O_j(x+t\hat{0},x_r+t\hat{0})]^{\dagger}_{cd} \, U^{\dagger}_{da}(x,x+t\hat{0}) \rangle \,.
\ees
We denote the static potentials by $V_{\alpha}(r),\;\alpha=0,1,2,...$.
The corresponding eigenstates of the Hamiltonian are denoted by
$|\alpha\rangle$.
Taking the limit of infinite time extension of the lattice
$T\to\infty$, we obtain from the transfer matrix formalism
the following spectral representation of \eq{potcorrev} \cite{Francesco:PhD}
\bes\label{potcorr}
 C_{ij}(t,r) & = & \sum_{\alpha} 
 \langle j|\alpha\rangle \langle\alpha|i\rangle\,\rme^{-tV_{\alpha}(r)} \,.
\ees
For fixed separation $r$, we extract from $C(t,r)$ the potentials
$V_{\alpha}(r)$ using the variational method described in \sect{variation}.
\begin{table}
 \centerline{
 \begin{tabular}{|l|l|} \hline
  fields & smearing levels $m$ \\ \hline\hline
  $U^{(m)}$ (see \cite{smear:ape} with $\epsilon=1/4$) & 7,10,15 \\ \hline
  $\Phi_2^{(m)}$ (see \eq{smearophiggs2}) & 1,3,5,7,10,15 \\ \hline
 \end{tabular}}
\caption{Here, we list the smearing levels for the gauge and 
  Higgs fields used in the simulation with parameters
  $\beta=2.4,\;\kappa=0.2759,\;\lambda=0.7$. \label{t_smear}}
\end{table}

\subsection{Results at ${\bf \beta=2.4}$} 

In our first study
\cite{Knechtli:1998gf} we obtained the static potential
from a simulation at $\beta=2.2,\;\kappa=0.274,\;\lambda=0.5$ on
a $20^4$ lattice. We observed string breaking at a distance 
$r_b/a\,\approx\,5$. We decided then to study the
system with a better lattice resolution at $\beta=2.4$.
The results that we describe in the following
are obtained on a $32^4$ lattice for the parameter set
\bes\label{parameterb24}
 \beta\,=\,2.4\,, \quad \kappa\,=\,0.2759\,, \quad \lambda\,=\,0.7 \, .
\ees
The field basis is constructed according to \eq{mesoneig} and
\eq{potbasis} from smeared gauge ($N_{\rmU}=3$)
and Higgs fields, whose smearing
parameters are summarised in \tab{t_smear}. 
The parameters for the simulation were fixed after some trial runs.
In the matrix correlation function $C_{ij}(t,r)\;(i,j=1,2,...,12)$,
the time-like links are replaced by their one-link integrals
\cite{Parisi:1983hm}.
We collected a statistics of 800 measurements.
Autocorrelations in the measurements
are practically absent: the statistical errors, 
computed by a jackknife analysis remain constant
when we group the measurements in bins of length 1,2 or 4.
\begin{figure}[tb]
\hspace{0cm}
\vspace{-1.0cm}
\centerline{\psfig{file=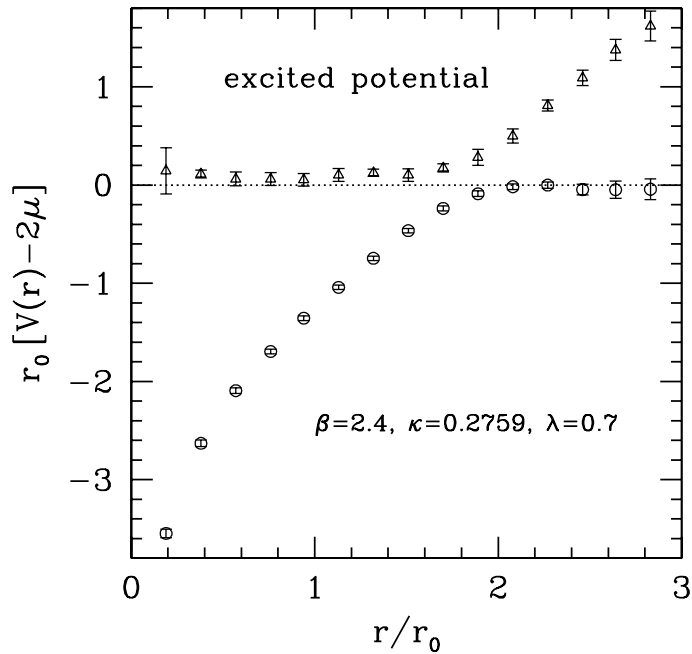,width=10cm}}
\vspace{-0.0cm}
\caption{Here, the renormalised ground state and first excited state static
  potentials in units of $\rnod$ are shown as functions
  of the separation of the
  static quarks. String breaking is clearly visible at
  $\rb\approx1.9\rnod$ together with the crossing of the energy
  levels. \label{f_invariant}}
\end{figure}

\subsubsection{Renormalised static potentials}

The static potentials $V_{\alpha}(r)\;(\alpha=0,1,2,...)$ are 
extracted from the matrix correlation \eq{potcorrev} using the
variational method described in \sect{variation}. We rewrite
\eq{spectrum} as
\bes\label{potentials}
   a V_{\alpha}(r) & = & \ln(\lambda_{\alpha}(t-a,t_0) 
  /\lambda_{\alpha}(t,t_0)) +
  \rmO\left(\rme^{-t\Delta V_{\alpha}(r)}\right) \, ,
\ees
where $\Delta V_{\alpha}(r)=
\min\limits_{\beta\neq\alpha}|V_{\alpha}(r)-V_{\beta}(r)|$ and the
eigenvalues $\lambda_{\alpha}(t,t_0)$ are obtained by solving the
generalised eigenvalue problem \eq{genev} with the matrix correlation
function at fixed $r$. We choose $t_0=0$. 
At all distances $r$ we can read off with
confidence and very good statistical precision (per mille level)
values for the static potential $V_0(r)$ at $t=7a$ which agree fully with $t=6a$. 
From the static potential along a lattice axis 
we determined the scale $\rnod$ exactly as explained in
\cite{Sommer:1993ce}. The result is
\bes\label{r0b24}
 \rnod/a & = & 5.29(6) \, .
\ees
Comparing this number with the values of $\rnod/a$ computed in
quenched QCD \cite{Guagnelli:1998ud}, we see that our point 
\eq{parameterb24}
in the SU(2) Higgs model corresponds in resolution to $\beta\approx6$ in
the SU(3) Yang-Mills theory with Wilson action. 
With respect to our first work \cite{Knechtli:1998gf}, the lattice
resolution is almost a factor 2 better.

The static potentials $V_{\alpha}(r)$ as they are obtained from \eq{potentials}
are not renormalised quantities because they contain self-energy contributions of the
static quarks which diverge like $1/a$ in the continuum.
We consider instead the differences $V_{\alpha}(r)-2\mu$ which are free of 
divergences
\cite{Francesco:PhD} and multiply them by $\rnod$ to obtain dimensionless
renormalised potentials.
In \fig{f_invariant}, we represent the ground state and the first
excited state static potentials.
The ground state potential shows an approximate
linear rise at intermediate distances: around separation
\bes\label{rb}
 \rb & \approx & 1.9\,\rnod
\ees
the potential flattens: the string breaks.
As expected, for large distances the potential approaches
the asymptotic value $2\mu$. The first excited potential comes
very close to the ground state potential around $\rb$ and rises linearly at
larger distances. The scenario of string breaking as a level crossing
phenomenon \cite{Drummond:1998ar} is confirmed beautifully.
\begin{figure}[tb]
\hspace{0cm}
\vspace{-1.0cm}
\parbox{7.2cm}{
\centerline{\psfig{file=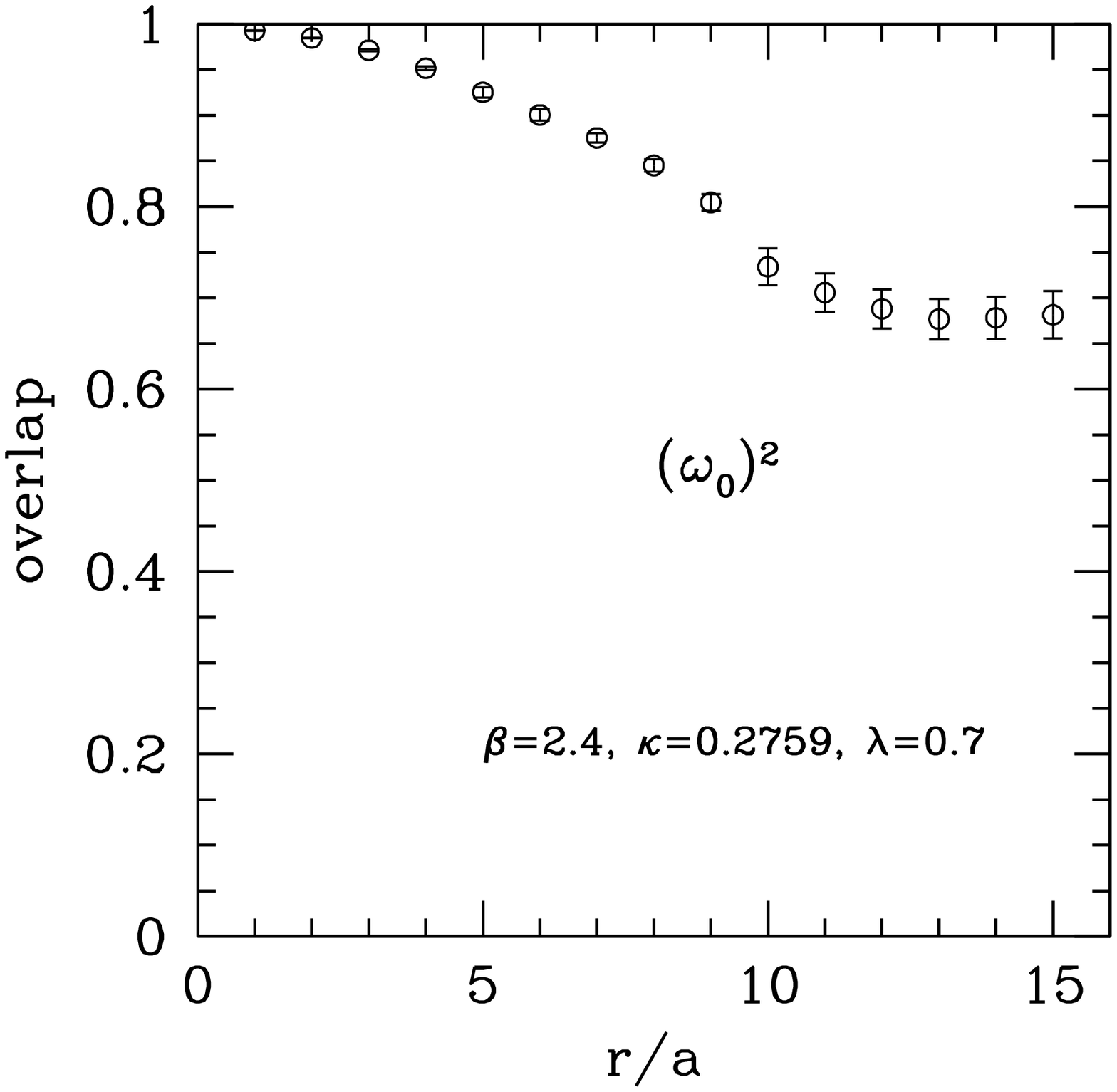,width=7.0cm}}}
\hfill
\parbox{7.2cm}{
\centerline{\psfig{file=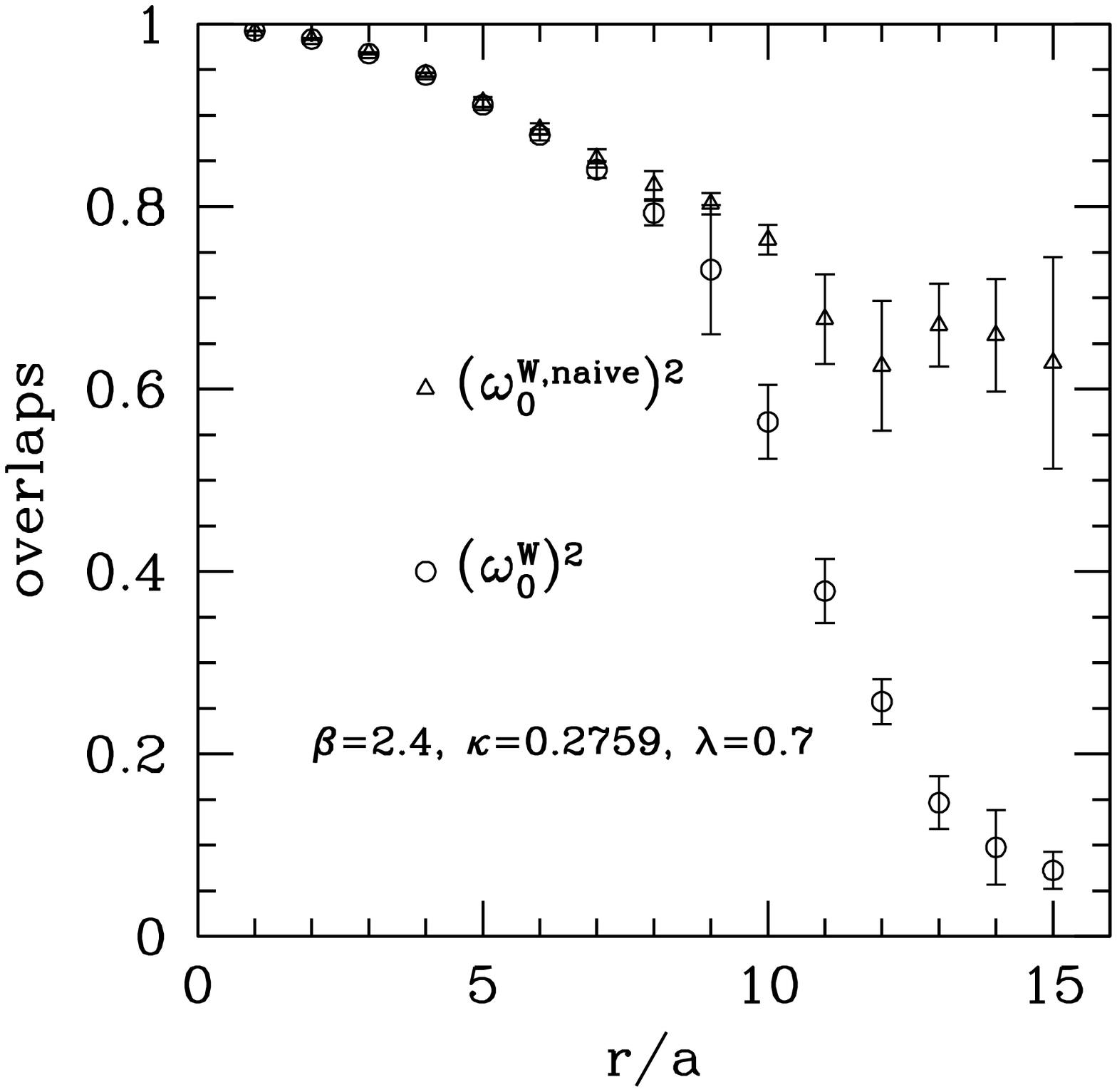,width=7.0cm}}}
\vspace{0.5cm}
\caption{In the figure on the left,
  we show the overlap $\omega_0$ obtained using
  the full basis of states \eq{potbasis}.
  In the figure on the right, we show the overlap
  $\omega_0^{\rm W}$ determined from string-type states only (described by the Wilson
  loops). A ``naive'' way of
  extracting it (triangles), using \eq{overlap}, gives an erroneous large overlap
  at long distances. A safe estimate (circles) is obtained from \eq{overlapwl}.
  \label{f_overlap0}}
\end{figure}

\subsubsection{Overlaps}
 
A certain measure for the efficiency of a basis of fields
\eq{potbasis} used to extract the ground state potential is given by
the {\em overlap}.
Using the approximate ground state wave function $v_0$ 
obtained from the variational method we 
define the projected correlation function
\bes\label{projcorr}
  \Omega(t) \; = \; \sum_{i,j}\,v_{0,i}C_{ij}(t)v_{0,j} \; = \; 
  \sum_{\alpha} (\omega_{\alpha})^2
  \rme^{-tV_{\alpha}(r)} \, , 
\ees
with normalisation $\Omega(t_0=0)=1$.
The positive coefficients $(\omega_{\alpha})^2$ 
may be interpreted as the square of
the overlap of the true
eigenstates of the Hamiltonian $|\alpha\rangle$ 
with the approximate ground state
characterized by $v_0$. The ``overlap'' is an abbreviation
commonly used to denote the ground state overlap, $\omega_0$.
We determine $\omega_0$ straightforwardly 
from the correlation function $\Omega(t)$ by noting that
\bes\label{overlap}
 \ln(\omega_0)^2 & \simttoinfty &
 \frac{t+a}{a}\ln\Omega(t)-\frac{t}{a}\ln\Omega(t+a)
  \, .
\ees                                                                            
We extract safe values for $(\omega_0)^2$
at $t=7a$, which agree fully with $t=6a$
and are shown in the left part of \fig{f_overlap0}.
Our basis of fields is big (and good) enough such that  
$(\omega_0)^2$ exceeds about 60\% for all distances.

It is interesting to consider also
the overlap for the (smeared) Wilson loops alone, i.e.
we restrict the matrix correlation function to the
$3\times3$ sub-block associated with string-type states. 
Let us denote the corresponding projected correlation function by
$\Omega_{\rm W}(t)$ and the overlap by $\omega_0^{\rm W}$. 
The computation of $\omega_0^{\rm W}$ is more difficult
because it turns out to be very small at large $r$. 
In the right part of \fig{f_overlap0}, we present the results for two estimates of
$(\omega_0^{\rm W})^2$.
The triangles correspond to the estimates $(\omega_0^{\rm W,naive})^2$
obtained directly from \eq{overlap},
with $\Omega(t)$ replaced by $\Omega_{\rm W}(t)$.
The circles correspond to the more reliable estimate using the
information from the full matrix correlation:
the expression
\bes\label{overlapwl}
  (\omega_0^{\rm W})^2 & \simttoinfty & (\omega_0)^2
  {\Omega_{\rm W}(t) \over \Omega(t)}
\ees
converges reasonably fast and $(\omega_0^{\rm W})^2$ can be estimated
from the r.h.s. for large $t$ ($t/a=7-9$ in practice).
Using \eq{overlapwl}, we see that
(smeared) Wilson loops alone have an overlap which drops at 
intermediate distances and
they are clearly inadequate to extract the ground state at large $r$.
On the contrary, $(\omega_0^{\rm W,naive})^2$ is
above 50\% at large distances: what is estimated here, is
actually the coefficient $(\omega_1^{\rm W})^2$, i.e. the square of the overlap of the
(smeared) Wilson loops with the first excited state (this statement
is supported by direct calculation, see \sect{s_mixing}).
The fact that $\omega_1^{\rm W}$ is so large
might explain the problems encountered in QCD
for observing string breaking from the analysis of a correlation matrix
with Wilson loops only.
\begin{figure}[tb]
\hspace{0cm}
\vspace{-1.0cm}
\parbox{7.2cm}{
\centerline{\psfig{file=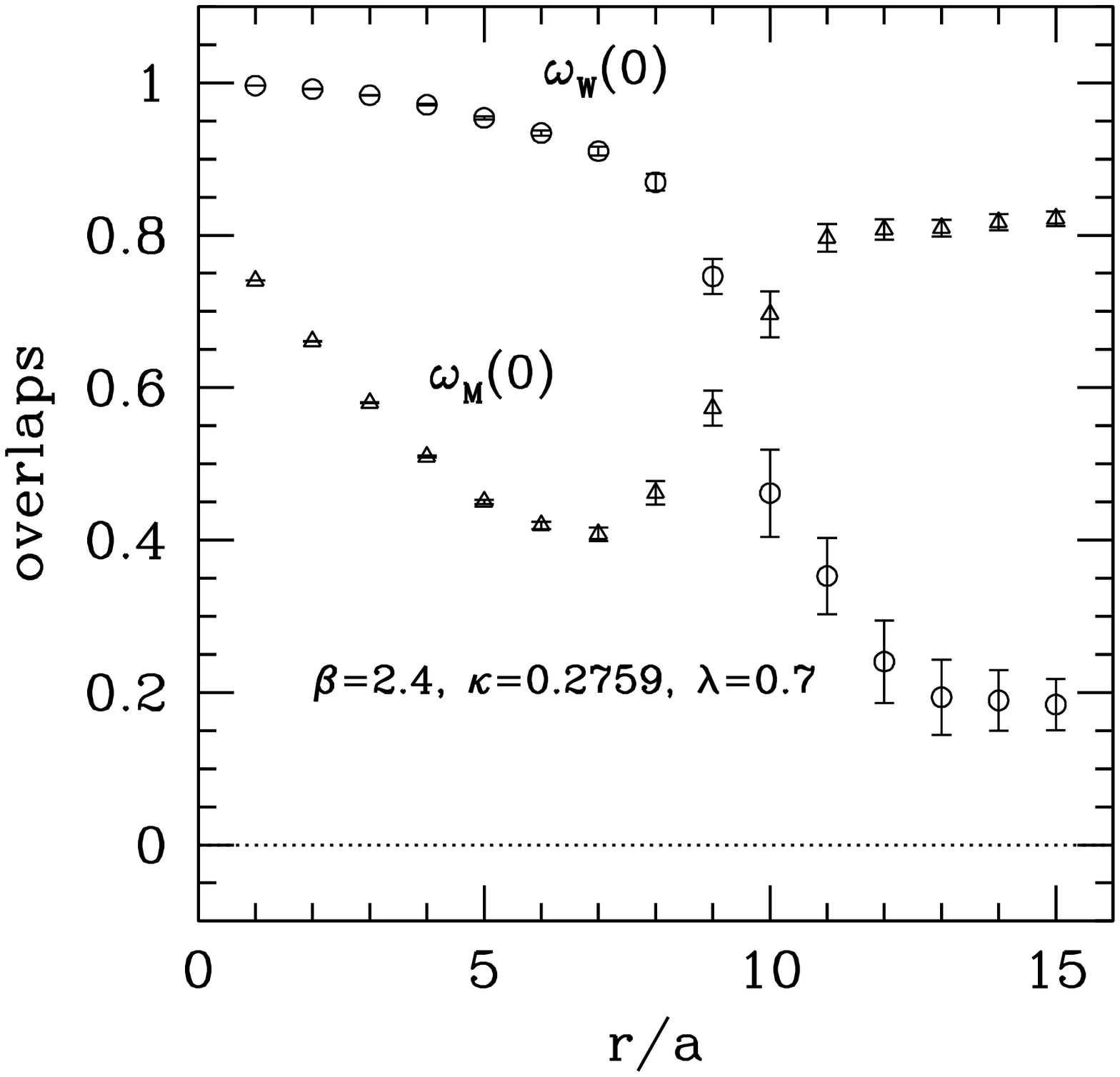,width=7.0cm}}}
\hfill
\parbox{7.2cm}{
\centerline{\psfig{file=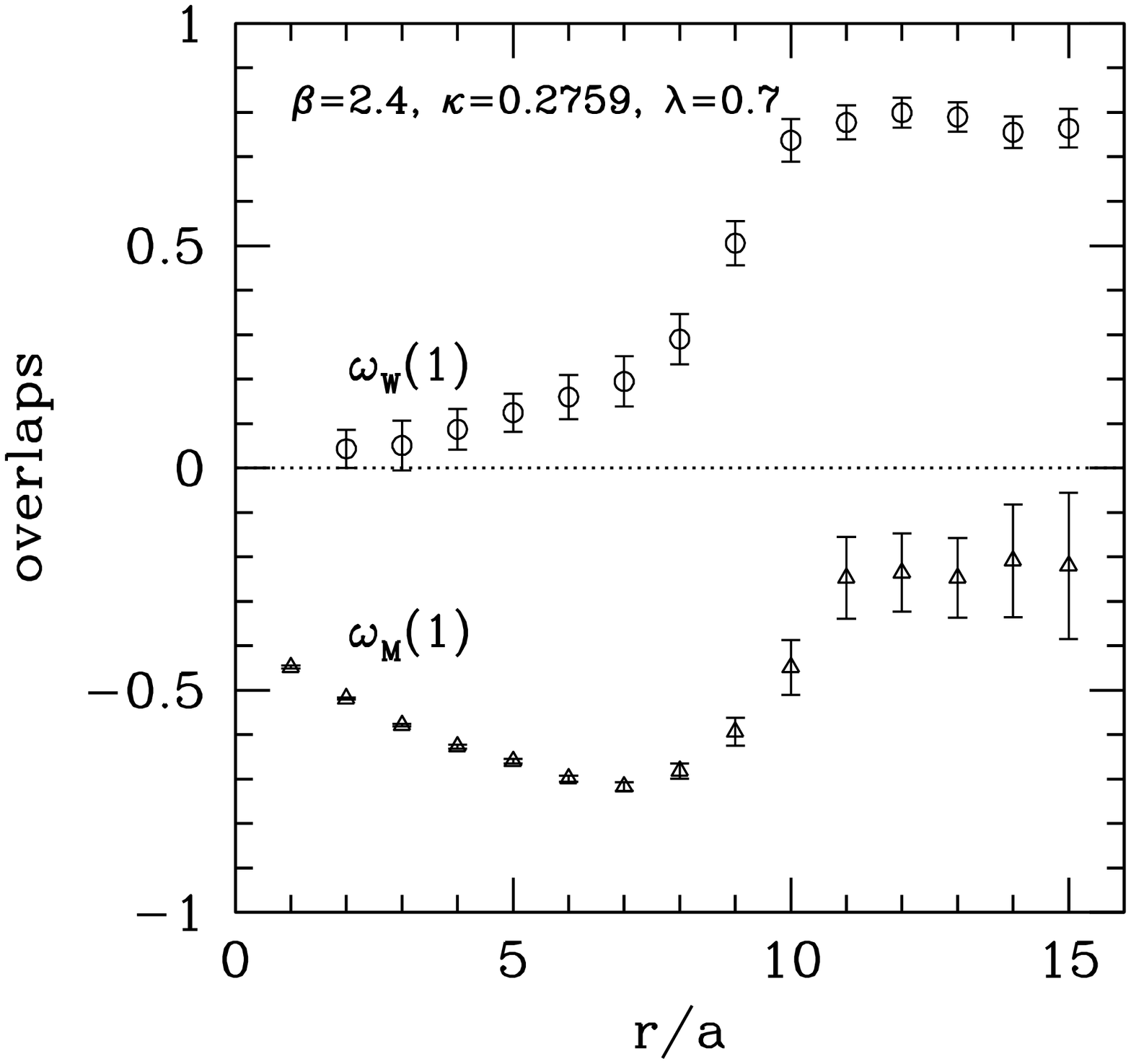,width=7.0cm}}}
\vspace{0.5cm}
\caption{In the figure on the left,
 the overlaps of the string-type (circles) and meson-type
 (triangles) states
 with the ground state of the Hamiltonian are shown
 as functions of the separation $r$ of
 the static quarks.
 In the figure on the right,
 the same overlaps but
 with the first excited eigenstate of the Hamiltonian are shown.
 \label{f_overlaps01}}
\end{figure}

\subsubsection{Mixing \label{s_mixing}}

Finally, we want to show that string breaking is
a mixing phenomenon involving string-type and meson-type states.
This leads to the crossing of the energy levels seen above.

We consider the diagonal sub-blocks of the matrix correlation
function \eq{potcorrev}
corresponding to string-type states (fields $i=1,2,3$ in
\eq{potbasis}) and to meson-type states (fields $i=4,5,...,12$ in
\eq{potbasis}) separately.
We determine approximate ground state wave functions $v_0^{\rmW}$ 
in the subspace of the string-type states
and $v_0^{\rmM}$ in the subspace of the meson-type states. With
the help of these wave functions we construct a $2\times2$ projected matrix
correlation function
\bes\label{projcorrwm}
 \Omega_{kl}(t)\,=\,\sum_{i,j}\,v_{0,i}^kC_{ij}(t)v_{0,j}^l\,=\,
 \sum_{\alpha} \langle\psi_l|\alpha\rangle
 \langle\alpha|\psi_k\rangle \rme^{-tV_{\alpha}(r)} \quad
 (k,l=\rmW,\rmM)\,,
\ees
where
\bes\label{smstates}
 |\psi_k\rangle\;=\;\sum_iv_{0,i}^k\,|i\rangle \,.
\ees
An inspection of 
\bes
  \Omega_{\rmW\rmM}(t_0=0)=\langle\psi_{\rmM}|\psi_{\rmW}\rangle
\ees
shows
that string-type and meson-type states are
orthogonal only for large values of $r$ \cite{Francesco:PhD}.
The coefficients
\bes\label{smoverlaps}
 \omega_k(\alpha) & \equiv & \langle\alpha|\psi_k\rangle \quad
 (k=\rmW,\rmM) \,,
\ees
in the expansion \eq{projcorrwm},
express the overlap of the string-type ($k=\rmW$)
and meson-type ($k=\rmM$) states with the
true eigenstates of the Hamiltonian.
We can choose our phase conventions for the 
states such that the coefficients
$\omega_k(\alpha)$ are real and in addition 
$\omega_{\rmW}(0)>0$ and $\omega_{\rmW}(1)>0$.
We truncate the sum in \eq{projcorrwm}
after $\alpha=1$ and consider the diagonal matrix elements
$\Omega_{kk}(t)$ for two fixed times $t=t_1$ and $t=t_2$: inserting the
known values for $V_0(r)$ and $V_1(r)$, we can solve
for $\omega_k^2(0)$ and $\omega_k^2(1)$.
The sign of the coefficients
$\omega_{\rmM}(0)$ and $\omega_{\rmM}(1)$ is fixed by the
off-diagonal matrix elements $\Omega_{\rmW\rmM}(t_1)$ and $\Omega_{\rmW\rmM}(t_2)$:
we find that for all $r$, $\omega_{\rmM}(0)>0$ and $\omega_{\rmM}(1)<0$ (in our sign convention).
The overlaps $\omega_{\rmW}(0)$ of the
string-type states (circles) and $\omega_{\rmM}(0)$ of the
meson-type states (triangles) with the
ground state of the Hamiltonian are shown on the left of \fig{f_overlaps01} and
the corresponding overlaps $\omega_{\rmW}(1)$ and $\omega_{\rmM}(1)$
with the first excited eigenstate of the Hamiltonian
on the right of \fig{f_overlaps01}.
String-type states have a large overlap at short distances
with the ground state and at large distances
with the first excited state.
Meson-type states have a large overlap at short
distances with the first excited state and at large distances
with the ground state.

In addition, we observe that the overlap
of the meson-type states with the ground state is also large at very short
distances. The explanation for this fact
is that string-type and meson-type states have an
overlap with each other at short distances.
In the string breaking region around $r/a=9-10$, the
overlaps of the string-type and meson-type states have similar
magnitude, both when the ground state or the first excited state is
considered. This fact is reflected in the crossing
of the energy levels \fig{f_invariant}.
Here, we would like to point out that the overlaps represented in
\fig{f_overlaps01} are not quantities which 
have a strict continuum limit. They depend on the $\beta$-value and the
other parameters (e.g. of the smearing) that we consider. However, as long as
one chooses a good basis (say with $\omega_0 > 0.5$) which can be separated 
into ``string like'' and 
``meson like'', the qualitative behavior in \fig{f_overlaps01} is expected to
persist also at smaller lattice spacings.


%% file: sect4.tex
\section{Scaling of the static potentials \label{s_scaling}}

\begin{figure}[tb]
\hspace{0cm}
\vspace{-1.0cm}
\centerline{\psfig{file=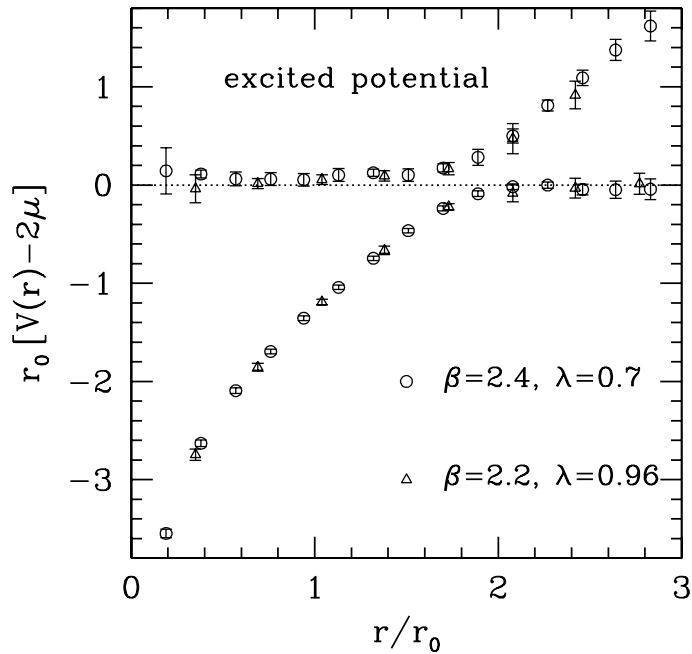,width=10cm}}
\vspace{-0.0cm}
\caption{Here, we show the scaling of the renormalised ground state and 
  first excited state static potentials. The parameter points lie on a line
  of constant physics.
  \label{f_invariant_lcp}}
\end{figure}
\begin{figure}[tb]
\hspace{0cm}
\vspace{-1.0cm}
\centerline{\psfig{file=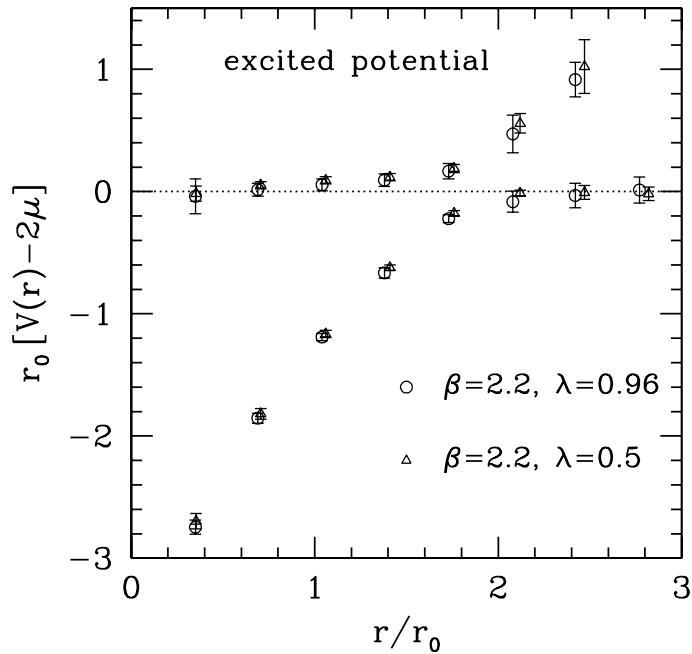,width=10cm}}
\vspace{-0.0cm}
\caption{Here, we show the $\lambda$-independence of the static
  potentials. The parameter $\kappa$ is determined from \eq{fit}.
  \label{f_invariant_b22}}
\end{figure}
In order to compare the renormalised static potentials at
different values of the lattice spacing
and estimate the size of scaling violations,
a way of determining {\em lines of constant physics} (LCP)
in the confinement ``phase'' of the SU(2) Higgs model is needed.
This question has been addressed in \cite{Knechtli:1999qe}, where a
non-perturbative determination of the LCPs is described.

The bare parameters $\kappa$ and $\lambda$ are renormalised along a LCP by 
keeping two
physical quantities $F_1$ and $F_2$ constant. A good choice is to take
\bes\label{F1}
  F_1 & = & \rnod\,[2\mu-V_0(\rnod)]
\ees                                                                            
and $F_2$ to be the generalised Binder cumulant $c_3$ defined in
\cite{Knechtli:1999qe}. \footnote{
Physically $F_1$ is a (non-perturbative) measure of the Higgs mass:
the variation of $F_1$ with the bare parameters is dominantly
caused by the variation of the meson mass $\mu$.
This bound state mass is of course expected to depend strongly
on the mass of its constituents. On the other hand, the interpretation of $F_2$
is less obvious. It was chosen to have a second renormalised
quantity which is both sensitive to the bare coupling $\lambda$ 
and can be computed in the MC simulations\cite{Knechtli:1999qe}.}

Taking the parameter set (\ref{parameterb24}) 
which we used in the simulation at $\beta=2.4$ we obtain the conditions
\bes\label{LCP}
  F_1 \, = \, F_1^*\,\equiv\,1.26 & \mbox{and} & F_2 \, = \, F_2^*
\ees
defining a LCP (the numerical value $F_2^*$ can be found in \cite{Knechtli:1999qe}).
As a result of the non-perturbative matching,
the parameter sets (\ref{parameterb24}) and
\bes\label{parameterb22}
 \beta=2.2\,, & \kappa^*=\kappa(\lambda^*)\,, & \lambda^*=0.96(10)
\ees
lie on a LCP. The value of $\kappa$ at $\beta=2.2$ is determined using the polynomial fit
\cite{Knechtli:1999qe}
\bes\label{fit}
 \kappa(\lambda) & = & 0.3131 + 0.0564\,(\lambda-1) - 0.0286\,(\lambda-1)^2
\nonumber \\ & &
+\, 0.0198\,(\lambda-1)^3 - 0.0246\,(\lambda-1)^4 
\ees
which is obtained by requiring $F_1=F_1^*\equiv1.26$
and correlates the uncertainties in $\kappa$ and $\lambda$ of the LCP.
In \fig{f_invariant_lcp}, we compare the results for the renormalised
ground state and first excited state static potentials that we
obtain at $\beta=2.4$ and at $\beta=2.2$ along the LCP.
The results for the potentials are compatible with scaling within minute 
errors under variation of the lattice spacing by almost a factor 2.

Another interesting issue is the $\lambda$-dependence of physical observables.
Exploratory results \cite{Montvay:1986nk} indicated that the physics of
the SU(2) Higgs model in the confinement ``phase'' is weakly dependent
on $\lambda$. This is certainly true near the continuum limit because it is
well accepted that the scalar part of the SU(2) Higgs model is a trivial
theory. Nevertheless, at finite value of the lattice spacing the model can be
considered as an {\em effective} field theory with three independent
renormalised couplings.
In \fig{f_invariant_b22}, we compare the static potentials for two
different values $\lambda=0.5$ and $\lambda=0.96$ at $\beta=2.2$.
The parameter $\kappa$ is determined from \eq{fit}.
There is no significant difference between the different $\lambda$ values.
This also means that the uncertainty in $\lambda^*$ in (\ref{parameterb22})
is irrelevant for our scaling test \fig{f_invariant_lcp}.


%% file: concl.tex
\section{Conclusions \label{s_concl}}

We presented the results for the static potentials (ground state and first
excited state) in the confinement ``phase'' of the SU(2) Higgs model. The
string breaking or flattening of the ground state potential
is clearly visible around separation $r_b\approx1.9\rnod$.
The comparison with the first excited potential shows a nice
crossing of the energy levels.
The interpretation of string breaking as level crossing phenomenon 
between string-type and meson-type states is
substantiated by the investigation of properly defined overlaps.

We also addressed the question of scaling violations in the measurements of the
static potentials. They are shown to be tiny already at $\rnod/a > 2.5$. 
Moreover, the dependence of the static potentials on the Higgs quartic
coupling $\lambda$ is very weak once the parameter $\kappa$ is determined by
keeping the physical quantity $F_1$ constant. 
These results are a strong indication for a continuum-like behavior of the
static potentials already at the relatively large values of the
lattice spacing that we used.

The method for the determination of the static potential that we presented
can be applied in QCD. Some steps in this direction have already been
made\cite{Pennanen:2000yk}. The main problem for QCD is the statistical accuracy. 
%

{\bf Acknowledgement.} We thank the Konrad-Zuse-Zentrum {f{\"u}r} Informationstechnik
Berlin (ZIB) for granting CPU-resources to this project.
